\documentclass[12pt,preprint]{aastex}

\usepackage{graphicx}
\usepackage{graphics}
\usepackage{natbib}

\newcommand{\be}{\begin{equation}}
\newcommand{\ee}{\end{equation}}
\newcommand{\nn}{\mbox{} \nonumber \\ \mbox{} }
\newcommand{\ba}{\begin{eqnarray}}
\newcommand{\ea}{\end{eqnarray}}
\newcommand{\om}{\omega}

\newcommand\etal{\textit{et al.\ }}
\newcommand\eg{\textit{e.g.,}}
\newcommand\cf{\textit{cf.\ }}

\newcommand{\Bf}{{magnetic field\,}}
\newcommand{\Bfs}{{magnetic fields\,}}
\newcommand{\Ef}{{electric  field\,}}
\newcommand{\Efs}{{electric fields\,}}

\newcommand{\NS}{neutron star\,}

\newcommand{\LC}{light cylinder}

\newcommand{\ms}{magnetosphere}

\newcommand{\Fermi}{{\it Fermi}}
\begin{document}

\title{The very-high energy emission from pulsars: a case for  inverse Compton scattering}
\author{Maxim Lyutikov\\
Department of Physics, Purdue University, \\
 525 Northwestern Avenue,
West Lafayette, IN
47907-2036 }

\author{Nepomuk Otte\\ 
Santa Cruz Institute for Particle Physics and Department of Physics,\\
University of California,
Santa Cruz, CA 
95060}

\author{
Andrew McCann\\
Department of Physics, McGill University,
Montr\'eal, QC, Canada  H3A 2T8}

\begin{abstract}

The observations of gamma-ray emission from pulsars with the \Fermi-LAT detector and the detection of  the Crab pulsar with the VERITAS array of Cherenkov telescopes at energies above 100  GeV make it unlikely that curvature radiation is the main source of  photons above GeV energies in the Crab and many other pulsars. 
We outline a model in which  the  broad UV-$X$-ray component  and the very high energy $\gamma$-ray emission of  pulsars are explained within the  Synchrotron-Self-Compton (SSC) framework. We argue that the bulk of the observed radiation is generated  by  the secondary  plasma, which is produced in cascades in the outer gaps of the magnetosphere. We find that the inverse-Compton (IC) scattering occurs in the Klein-Nishina regime, which favors synchrotron photons in the UV band as target field for the scattering process. 
The  primary beam is  accelerated in a modest electric field, with  a field strength that is of the order of a few percent of the \Bf\ near the light cylinder.   Overall,  in the Klein-Nishina regime of the IC scattering  the  particle distribution in the gap does not evolve towards a stationary  distribution and thus  is intrinsically time-dependent.  
We point out that in a   radiation reaction-limited regime of particle acceleration the gamma-ray luminosity $L_\gamma $  scales {\it linearly}  with the pulsar spin-down power $\dot{E}$, $L_\gamma \propto \dot{E}$, and not proportional to $\sqrt{\dot{E}}$ as expected from potential-limited acceleration.

\end{abstract}

\section{Introduction}

The recent launch of the \Fermi\ Gamma-Ray Space Telescope and subsequent detection of a large number of pulsars \citep{2010ApJS..187..460A} revolutionized our picture of the non-thermal emission from pulsars in the gamma-ray band from 100\,MeV up to about 10\,GeV. At even higher energies, in the very-high energy (VHE) band, the detection of the Crab pulsar at 25\,GeV by the Magic Collaboration  \citep{2008Sci...322.1221A} and recently at 120\,GeV by the VERITAS Collaboration  \citep{VERITASPSRDetection} in the very-high energy (VHE)  band  allow to stringently constrain the very-high-energy emission mechanisms in the case of the Crab pulsar. In this paper we show that it is very difficult to invoke curvature radiation as the dominant radiation mechanism to explain the observed emission above 100\,GeV and, furthermore, demonstrate  that inverse-Compton (IC) upscattering of UV photons into the VHE band can explain the observations in the gamma-ray band.

\cite{1986ApJ...300..500C} were amongst the first to discuss high energy emission from the magnetosphere of pulsars. They proposed the outer gap as the location where charged particles accelerate to relativistic energies and radiate in the gamma ray band. The outer gap model is currently one of the most favored models to explain non-thermal radiation from pulsars. \footnote{Below, for the order-of-magnitude estimates, by ``outer gaps models''  we imply   generic ``outer magnetosphere models''.} Based on the idea of the outer gap, geometrical models are very successful in explaining the basic features of the observed $\gamma$-ray light curves \cite[\eg][]{1995ApJ...438..314R,2008ApJ...680.1378H,2010ApJ...715.1282B}. While there seems broad consensus that the particle accelerator is located in the outer magnetosphere, the radiation physics remain controversial. One of the preferred radiation mechanisms, which is believed to dominate the observed gamma-ray emission, is curvature radiation \cite{1996ApJ...470..469R} (see also \cite{1986ApJ...300..500C,2000ApJ...537..964C,2008MNRAS.386..748T,2008ApJ...676..562T}). Possible importance of the IC scattering was mentioned previously \citep[\eg][]{1996ApJ...470..469R}, but  was never considered the primary emission mechanism for the very high energy photons. 

This paper is structured in the following way. In \S \ref{curv} we demonstrate that the recent  results obtained with \Fermi\  \citep{2010ApJS..187..460A}, Magic \citep{2008Sci...322.1221A}  and especially VERITAS \citep{VERITASPSRDetection} make it highly unlikely that curvature emission is the main radiation mechanism of photons above 10 GeV energies from the Crab pulsar.  In  \S \ref{IC} we show that inverse-Compton scattering by secondary  particles in the outer gaps is broadly consistent with  the observed luminosity in the very-high-energy band.

\section{Limits on curvature radiation}
\label{curv}

\subsection{Crab  pulsar}

Curvature radiation is a widely discussed process to explain the observed gamma-ray emission from the magnetosphere of pulsars. In this section we discuss the difficulty of invoking curvature radiation as the emission process that explains the observed pulsed emission from the Crab pulsar above 100 GeV. 

We assume an outer gap scenario with the accelerating electric field  being parallel to the magnetic field \cite{1986ApJ...300..500C}. In the electric field a beam of charged particles accelerates -- hereafter primary beam -- that has a particle density, which is of the order of the Goldreich-Julian density $n_{GJ}$ \citep{GoldreichJulian}. The primary beam loses a significant amount of its energy through various radiative processes of which the curvature emission and the  IC-induced  pair production are the dominant ones \cite{1983ApJ...266..215A,cr77}. The pair-production process results in the formation of a second population of particles -- hereafter secondary plasma --, which has a higher particle density than the primary beam but a smaller bulk Lorentz factor. Within this outer-gap framework we derive a general upper limit of the break in the curvature radiation spectrum that is emitted by particles within the outer gap of the Crab pulsar. The limit we obtain is independent of the particular details of the acceleration mechanism of the primary beam. In our argument we follow a similar approach that has been applied before in the discussion of the synchrotron emission from pulsar wind nebulae by \cite{1996ApJ...457..253D} and \cite{2010MNRAS.405.1809L}. 

Within the outer gap, the charged particles follow the curved magnetic field lines and, therefore, emit curvature-radiation photons. The curvature radiation spectrum emitted by monoenergetic particles has a break at energy $\epsilon_{br}$ \citep{1996ASSL..204.....Z}
\be
\epsilon_{br} = { 3 \over 2} \hbar  { c \over R_c} \gamma_b^3,
\label{00}
\ee
where $R_c$ is the curvature radius of the magnetic field lines, and $\gamma_b$ is the Lorentz factor of the radiating particles. 

An upper limit of $\gamma_b$ is set by the constraint that while the particles accelerate they radiate and, therefore, the maximum value of $\gamma_b$ is obtained when acceleration gains are balanced by radiative losses, i.e. the radiation reaction limit. Under the assumption that the accelerating electric field $E$ is a fraction $\eta \leq 1$ of the magnetic field $B$, the acceleration gain is $ec\eta B$, with $e$ is the electron charge. The radiation reaction limit is then reached if:
\be
e c \eta B = {2 \over 3}  { e^2 \over c} \gamma_b^4 \left( { c\over R_c} \right)^2,
\label{01}
\ee
where the losses due to curvature radiation are given on the right side. Using Eq.\ (\ref{01}) it follows from Eq.\ (\ref{00}) that 
\be
\epsilon_{br} = \left( { 3 \over 2} \right)^{7/4}  \hbar  c \sqrt{R_c} \left( \eta  { B \over e} \right)^{3/4} 
\ee
The radius of curvature $R_c$ can be expressed in units of the light cylinder $R_L$, $R_c = \xi R_L = \xi c P/ ( 2 \pi)$, where $P$ is the period of the pulsar and $\xi$ is a dimensionless scaling parameter. If, furthermore, $B$ is replaced by the radial distribution of the magnetic field of a dipole $B = B_{NS} (R_{NS}/R)^3$ \footnote{Very close to the \LC\ the toroidal \Bf\ induced by the poloidal currents becomes important; we neglect here these effects.} , where $B_{NS}$ is the magnetic field on the surface of the neutrons star and $R_{NS}$ the starÕs surface, then it follows that:
\ba
&&  
\epsilon_{br} = (3 \pi)^{7/4}  { \hbar \over ( c e) ^{3/4} } \eta^{3/4} \sqrt{\xi} \,  {  B_{NS}^{3/4} R_{NS}^{9/4}\over P^{7/4}} =
150\, {\rm GeV} \, \eta^{3/4} \sqrt{\xi}  = 5\, {\rm GeV} \, \eta_{-2} ^{3/4} \sqrt{\xi} 
\nn &&
\gamma_{b}  = (3 \pi)^{1/4}   { 1 \over ( c e) ^{1/4} } \eta^{1/4} \sqrt{\xi} \,  {  B_{NS}^{1/4} R_{NS}^{3/4}\over P^{1/4}} =9 \times 10^7 \,
 \eta^{1/4} \sqrt{\xi}  = 3 \times 10^7  \eta_{-2} ^{1/4} \sqrt{\xi}\,.
 \label{1}
\ea
On the very right  side we parametrized  $\eta = 10^{-2}  \eta_{-2}$,  assuming that the electric field is a few percent of the magnetic field. In  the current outer-gap models \citep{1986ApJ...300..500C,2000ApJ...537..964C,2008MNRAS.386..748T,2008ApJ...676..562T}   
an electric field of $E _\parallel \approx ( \Omega B r^2)/(c R_c)\sim 0.1 B$ is predicted, while in the models of \cite{2000ApJ...537..964C,2008MNRAS.386..748T,2008ApJ...676..562T} the accelerating field is one order of magnitude smaller.

In the radiation reaction limited regime, the maximum energies of photons emitted by curvature radiation is determined by the maximum energies of the electrons. The result is a break in the spectrum and an exponentially falling flux above the break. The gamma-ray spectrum of the Crab pulsar has a break at about 6 GeV \citep{2010ApJS..187..460A}, which is formally consistent with the break we predict for an electric field that is a few percent, i.e.\ $\eta _{-2} \sim 1$.  However, the observed gamma-ray flux above the break is not exponentially falling, which is expected if the break would be due to curvature radiation produced by the electrons with the highest energies. The non-exponential cutoff can be explained if the electric field is larger, i.e.\ $\eta\sim 1$, or if a different emission mechanism dominates above the break energy.

\subsection{Other pulsars}

\begin{figure}[htb]
% \vspace{-15pt}
\includegraphics[width=0.99\linewidth]{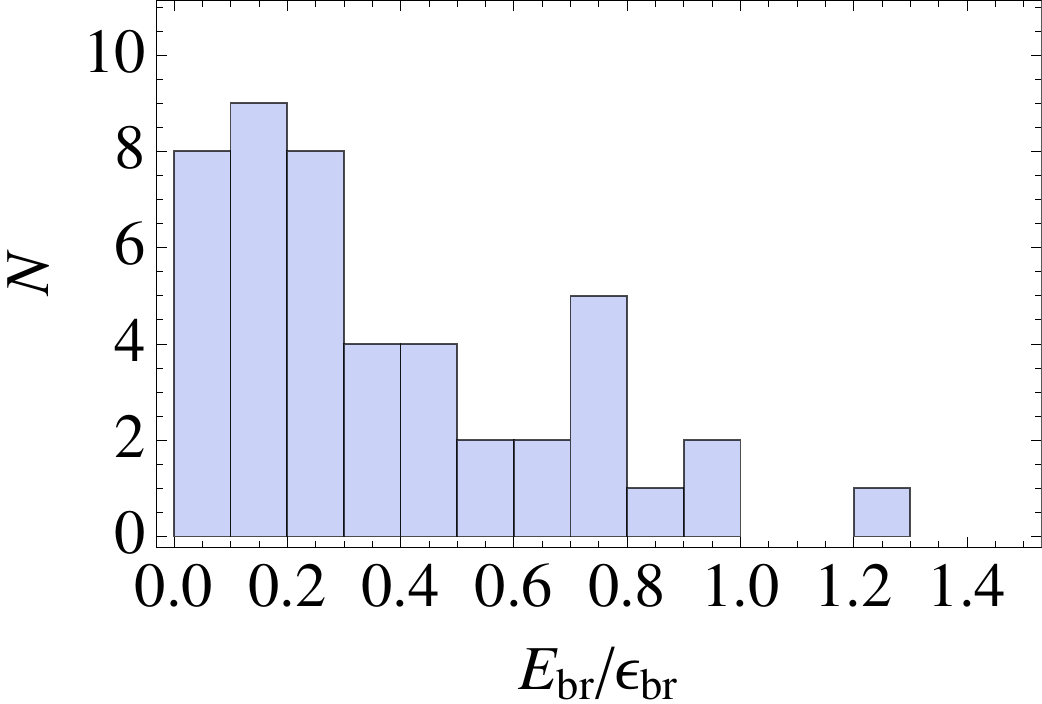}
% \vspace{-35pt}
\caption{Ratio of the  observed break energies $E_{br}$ for 46 pulsars to the maximum predicted for curvature radiation $\epsilon_{br}$, which is given by Eq.\ (\ref{1}) with $\eta =\zeta=1$}
\label{ratio}
\end{figure}

In the gamma-ray band a few dozen pulsars have been detected With the \Fermi-LAT detector. The spectral energy distributions of these pulsars in the $\gamma$-ray band are very similar and can be characterized by a flat spectral component between 100 MeV and a few GeV \citep{1999ApJ...516..297T,2010ApJS..187..460A} and a spectral break in the GeV region. In Fig.\ \ref{ratio} we compare the observed spectral breaks of these pulsars  with the predicted ones from Eq.\ (\ref{1}) by calculating the ratio of the observed $E_{br}$ and predicted spectral break $\epsilon_{br}$ for each of the 46 pulsars reported in the first \Fermi\ catalogue \citep{2010ApJS..187..460A}. For the calculation of the break energy $\epsilon_{br}$ we used the extreme case of $\eta =\zeta=1$. 
If the spectral break is due to curvature radiation and the electric field in the gap is much less than one, $\eta << 1$, as it is expected in present outer gap models, the ratio should be much smaller than one. This is indeed the case for the majority of the pulsars, including the Crab pulsar. However, for a significant number of pulsars the ratio is close to one and for one pulsar, PSR  J1836 + 5925, the ratio is even larger than one. In order to explain the spectral break for these pulsars as a result of curvature radiation an accelerating \Efs\ is required that is close to or even larger than \Bfs. 
 
Two possible interpretations of these results are: (i) the observed spectral break is due to curvature radiation by the electrons with the highest energies. For the Crab pulsar a new component dominates above the break and explains the non-exponential cutoff. In this interpretation, the pulsars, for which the ratio $E_{br}/\epsilon_{br}$ is close to unity, can be explained by statistical outliers (uncertainties on $E_{br}$ are not taken into account  in Fig. \ref{ratio}), or that a different emission mechanism dominates at high energies that influences the measurement of $E_{br}$. (ii) The gamma-ray emission above $\sim$ GeV energies is due to one single emission process, which is not curvature radiation. In this case the spectral break reflects the underlying particle  distribution.

\section{ IC model of the high-energy gamma-ray emission from the Crab pulsar}
\label{IC}

\subsection{Outline of the model}

In this section we outline the key features of an SSC model that is able to explain the  high energy emission of the Crab pulsar.  Observationally, 
the spectral energy distribution (SED) of the Crab pulsar has a broad peak in the 10-100 keV range with   a luminosity $L_X \approx 10^{36} {\rm erg s} ^{-1}$  \citep[cf.\ Fig.\ 9 in][]{2001A&A...378..918K} which is a few percent of the pulsar's spin-down power of $\approx 5\times 10^{38} {\rm erg s} ^{-1}$.  Between $10 $ MeV and the spectral break at a few GeV the SED is flat and has a luminosity $L_\gamma$ of a few $\times 10^{34}  {\rm erg s} ^{-1}$  \citep{2010ApJ...708.1254A}. Above the spectral break at $\geq 150 $ GeV the luminosity is $\sim 10^{33} $erg s $^{-1}$.

We identify the broad soft UV-$X$-ray peak in the SED of the Crab pulsar as a synchrotron (or possibly cyclotron) emission from the secondary plasma boosted by the large parallel velocities of emitting particles. This creates target photons for IC scattering both by the primary beam and by the  secondary plasma. As we demonstrate below, the IC scattering by the secondary plasma is broadly consistent with the observations. 

\subsection{IC scattering by the primary beam}
In this section we discuss the inverse Compton scattering by the primary beam in the outer gap. We use, like in the previous section,  $E =\eta B = 10^{-2} \eta_{-2} B$ for the accelerating \Ef\ and $\eta_ G R_{LC}^3$ with $\eta_ G = 0.1\,  \eta_ {G,-1}$ for the effective emitting volume. In order to simplify our calculations, we separate the broad UV-$X$-ray peak into two component: A low energy component that covers the  UV band with a luminosity of $L_{UV} \approx 10^{34} {\rm erg s} ^{-1} l_{34}$ and typical photon energies of $\epsilon_{\rm soft}=   1 \, {\rm eV}\,\epsilon_{UV,0}$, and a high energy component that is centered around the $X$-ray peak with a luminosity of $L_{X} \approx 10^{36} {\rm erg s} ^{-1}  l_{36}$ and  typical photon energies of $\epsilon_{\rm soft}=   1 \, {\rm keV}\,\epsilon_{X,3}$. The need to separate the broad-band component into two comes from the strong dependence of the IC scattering in the KN regime on the energy of the photon .

The properties of Inverse Compton scattering strongly depend on whether the scattering occurs in the Thompson regime or in the Klein-Nishina (KN) regime. In which regime the scattering takes place is determined by the Lorentz factor of the scattering particle and the energy of the upscattered photon \citep{1970RvMP...42..237B}. For a given photon energy $\epsilon$  the scattering takes place in the KN regime if the Lorentz factor $\gamma_{KN}$ is larger than 
\be 
\gamma_{KN} = {1 \over 4} { m_e c^2 \over \epsilon_{\rm soft}}\approx 1.2 \times 10^5\,  \epsilon_{UV,0}^{-1}  \approx 1.2 \times 10^2\,  \epsilon_{X,3}^{-1}
\label{gKN}
\ee
These are fairly modest Lorentz factors considering the above estimate of the maximum Lorentz factor that can be achieved in the outer gap in the radiation reaction limit, Eq. (\ref{1}).  It can, therefore, be concluded, that inverse Compton scattering by the primary beam takes place in the KN regime.

Adding losses due to inverse Compton scattering  in the extreme KN limit into the balanced gain loss equation (\ref{01}) results in a net energy loss of \citep{1970RvMP...42..237B,2010NJPh...12c3044S}
\be
 \dot{\epsilon}  = e c \eta B - {2 \over 3}  { e^2 \over c} \gamma^4 \left( { c\over R_c} \right)^2- {4 \over 3}  \left( {m_e c^2 \over \epsilon_{\rm soft}} \right)^2 U_{\rm soft} \sigma _T c ,
 \label{eKN1}
 \ee
where $U_{\rm soft}$ is the energy density of the target photon field, $\epsilon_{\rm soft}$ is the typical energy of a soft photon, and $\sigma_T$ is the Thompson cross-section.
Note that both the acceleration term and the decelerating  IC  term are independent of the energy of the particle. Thus, if curvature losses were negligible, particles are either accelerated or decelerated without reaching a steady solution. Only in the presence of curvature radiation is it possible to achieve a steady-state particle distribution.

In order to better understand how curvature radiation and inverse-Compton scattering contribute to the radiation loss of the primary beam in Eq. (\ref{eKN1}) we compare the two. In this comparison we assume the curvature radiation-limited Lorentz factor Eq.\ (\ref{1}), and justify  our choice {\em post factum} by showing that curvature radiation and IC losses in the Crab pulsar are about equal. The  soft photon luminosity that results in IC losses in the KN regime which are similar to curvature radiation losses is
  \be 
  L_{\rm soft, crit} = \eta { B_{NS} R_{NS}^3 \Omega  \epsilon_{UV}^2 \over e^3} = 
  \left\{
  \begin{array}{ll}
   10^{35} \,  {\rm erg s}^{-1} \, \epsilon_{UV,0}^{2}\eta_{-2}
  \\
   10^{41} \,  {\rm erg s}^{-1} \epsilon_{X,3}^{2}\eta_{-2}
  \end{array}
  \right.
  \label{Lcrit}
  \ee
The minimum luminosity of the target photon field in the UV that is needed to achieve IC losses similar to curvature radiation losses Eq.\ (\ref{Lcrit}) is about the same as the  observed UV luminosity. The upscattering of soft X-ray photons, even though the X-ray flux is higher than the UV flux,  does not contribute much to the radiative loss of the primary beam (KN suppression) because the observed X-ray luminosity is five orders of magnitude below the critical luminosity.
   
The conclusion that the IC upscattering of UV photons and curvature radiation contribute  about equal to the total loss of the primary beam means that both processes also contribute equally to the emitted power in the gamma-ray band. However, the two  processes produce very different spectral features. As we have shown before, curvature radiation photons can only be emitted with energies up to a few GeV for reasonable electric fields and curvature radii, see Eq. (\ref{1}). The spectrum of the IC upscattered photons, on the other hand, extends to much higher energies. This can be shown by assuming again that curvature radiation and IC losses are about equal, in which case the maximum Lorentz factor can still  be estimated with Eq. (\ref{1}). The maximum energy of the upscattered photons, $\epsilon_\gamma$,  is then given by the maximum electron energy: 
   \be
   \epsilon_\gamma \approx \gamma_b m_e c^2 = ( 3 \pi)^{1/4} { m_e c^{7/4} \over e  ^{1/4} } \eta^{1/4} \sqrt{\xi} \,  {  B_{NS}^{1/4} R_{NS}^{3/4}\over P^{1/4}} = 15 \, {\rm TeV} \, \eta_{-2} ^{1/4} \sqrt{\xi}
   \label{gg}
   \ee

While the maximum photon energy produced by IC scattering depends on the maximum electron energy, the total power emitted by IC scattering is independent of the electron energy. Instead the total power is determined by the low-energy target photons field. Due to the steeply falling IC cross-section in the KN regime with increasing energy of the target photons $\propto \epsilon^{-2}$, the maximum IC power $L_{KN}$ might not be determined by the peak luminosity in the spectral energy distribution of the target photons but be at lower energies: 
\be 
%what exactly is this, the peak luminosity, the integrated luminosity...
L_{KN,b} =  \left( {m_e c^2 \over \epsilon_{\rm soft }} \right)^2 U_{\rm soft} \sigma _T c  \times n_{GJ} \times  \eta_G R_{LC}^3 \label{LKN}
\ee
Application to the Crab pulsar yields that the primary beam produces an IC luminosity by upscattering the X-ray photons with keV energies, luminosity of $L_{X} \sim 10^{36} $ erg $s^{-1}$, that is,
\be 
%check numbers
L_{KN, X} =5 \times10^{29}\,   \eta_{ G,-1}   \, \epsilon_{X,3} ^{-2} 
\ee
This is much lower then the IC luminosity produced by upscattering the UV photons with eV energies, $L_{UV} \sim  10^{34} $  erg $s^{-1}$:
\be
% check numbers
L_{KN, UV } =5\times10^{33} \,   \eta_{ G,-1}\,  \epsilon_{UV,0} ^{-2}\, .
\label{OK}
\ee
The above is an estimate of the peak power. The average luminosity is lower by at least one order of magnitude. Thus, we conclude that the IC scattering by the primary beam is unlikely to be the origin of the VERITAS signal.

\subsection{Gamma-ray emission from the secondary plasma}

In the previous section we discussed the gamma-ray emission produced by the primary beam. In this section we discuss the gamma-ray emission by the particles that are produced in pair cascades of the particles in the primary beam, the secondary plasma.

We recall that the primary beam has a  density $n_{GJ}$ and a Lorentz factor $\gamma_b$ (\ref{1}). As nomenclature for the secondary plasma we use $n_p$ for its density and $\gamma_p$ for its Lorentz factor. We assume energy equipartition between the primary beam and the secondary plasma  \citep[the assumption of equipartition between the primary beam and secondary plasma is justified in the polar cap models;][ we assume a similar parametrization here]{1996ApJ...458..278D}.   From equipartition it follows that $n_p \gamma_p = n_{GJ} \gamma_b$. The two particle populations are connected through the pair cascading process, i.e.\ $n_p = \lambda_p n_{GJ}$, where $\lambda = 100 \lambda_2 $ is the multiplicity factor of the secondary particles.  Multiplicities of the order $\lambda \sim 10^2$ are typical in outer gap models \citep[\eg][]{2011ApJ...736..127W}, but can  also reach much higher values, $\lambda \sim 10^4-10^6$ \citep{2010ApJ...715.1318T}.

In our picture of a radiation-reaction limited acceleration of the primary beam it follows that the Lorentz factor  of the secondary plasma is given by
\be
\gamma_p \approx  
\gamma_b/\lambda  =  3 \times 10^5  \eta_{-2} ^{1/4} \sqrt{\xi} \,   \lambda_{2}^{-1}
\label{gammap}
\ee
This Lorentz factor is above the minimum $\gamma_{KN}$. Therefore, IC  scattering by the secondary plasma takes place in the KN regime and we can use the same relations that we have derived in the previous section for the emission produced by the primary beam. 
(For multiplicities mush higher than the assumed $\lambda  \approx 100$ the scattering by UV photons occurs in the Thompson regime. Overall, the convolution of the electron and the soft photon spectrum requires detailed radiative calculations which include global magnetospheric models and anisotropic angular distributions of the  photons.) The maximum energy of  IC photons produced by the  secondary  plasma is (\cf\ Eq. (\ref{gg}))
 \be
   \epsilon_{\gamma , p} \approx \gamma_p  m_e c^2 = 150 \, {\rm GeV} \, \eta_{-2} ^{1/4} \sqrt{\xi}\,  \lambda_{2}^{-1}
   \label{22}
   \ee
 and the peak luminosity of the IC scattered UV photons is  (\cf\ Eq. (\ref{OK}) is
   \be
   L_{KN,p  } =\lambda L_{KN,b} = 4\times10^{35}   \eta_ {G,-1}  \epsilon_{UV,0} ^{-2}   \lambda_{2} 
\label{OK1}
\ee
Both the  energy  (\ref{22}) and the peak  luminosity (\ref{OK1}) are consistent with the VERITAS detection.
% Check!
 Thus, IC up-scattering of UV photons by the secondary plasma can explain the observed pulsed emission from the Crab pulsar above 100 GeV. We leave a more detailed calculation of the spectrum to a future paper (we expect that the overall spectrum will depend on the distributions both in parallel $p_\parallel$ and perpendicular  $p_\perp$ momenta).

The secondary plasma also produces synchrotron photons with energies that can be estimated, \eg\ using Doppler-boosted cyclotron emission:
\be
  \epsilon_{X,p} = \hbar \om_B \gamma_p = \eta^{1/4} \sqrt{\xi}  { \hbar  e^{3/4} B_{NS}^{5/4} \Omega^{13/4} \over \lambda m_e c^{17/4}}=
  3\, {\rm keV} \eta_{-2}^{1/4} \sqrt{\xi} \lambda_{2}^{-1}\,. 
  \ee
  This roughly coincides with the energy where the Crab pulsar emits most of its power (in fact, the Crab emits most of it's power around 100 keV). 
  
  To produce the observed synchrotron luminosity $L_s \approx  N_p (e^2/c) \om_B^2 \gamma_\perp^2 \gamma_p^3 \approx 10^{36} {\rm erg s} ^{-1}$  ($N_p$ is the total number of secondary particles in the \ms, $\gamma_\perp$ is a typical transverse Lorentz factor) one requires
  \ba &&
  N \approx 5 \times 10^{32} \epsilon_{UV, 0}^{-1}  \gamma_{p,2}^{-2}
  \nn &&
  \gamma_\perp= { \sqrt{ \epsilon_{UV, 0}} \over \sqrt{\gamma_{p,2}}}
  \ea
  This demonstrates that for the chosen parameters $\gamma_\perp \sim 1$, the soft emission occurs in the cyclotron regime, and it also shows that the overdensity
  \be
  \lambda = {N /R_{LC}^3 \over n_{GJ}(R_{LC}) } = 60  \gamma_{p,2}^{-2}  \epsilon_{UV, 0}^{-1} , 
  \ee
  is consistent with our assumption of $\lambda _2 \sim 1$.
  
We, therefore, conclude that emission from the secondary plasma is not only able to explain the observed gamma-ray emission above 100 GeV by upscattering UV photons but it also explains the bulk of the $X$-ray emission. An obvious modification is required to this simplified picture to include the relativistic momenta of the secondary particles that is transverse to the magnetic field lines and results in synchrotron and not cyclotron emission as we assumed. 
%How much does it change? Can we give an estimate to understand the uncertainty/validity of the assumption we make?

\section{Expected $X$-ray-$\gamma$-ray correlations}

Within the framework of the SSC model the power emitted by IC  is related to the power of the seed photons. Photons of different energies that are emitted by the same particles should in principle produce similar pulse profiles. In our model one expects, therefore, that the pulse profiles in X-ray and in gamma-rays are similar because the secondary plasma emits synchrotron radiation in X-rays and IC scatters UV photons into the VHE band. And indeed, the ratio of the amplitudes of the two pulses in the pulse profile of the Crab pulsar changes consistently in the X-rays / soft gamma-ray band and in the high energy gamma-ray band. In X-rays the main pulse dominates over the inter pulse. The ratio changes towards higher energies and reverses in the soft gamma-ray band at about 1\,MeV. Similarly, the main pulse dominates at 100\,MeV  \citep[see, \eg also for the pulse profiles at lower energies][]{2010ApJS..187..460A} while at 120\,GeV the inter pulse clearly dominates over the main pulse \cite{VERITASPSRDetection}.

In addition, as  we have argued that though IC losses  may be energetically dominant (or similar to curvature emission), in the KN regime they do not lead to a equilibrium distribution of Lorentz factors. Hence we expect highly non-stationary magnetospheric plasma flows. This will lead to highly non-stationary radiative properties. Since within the SSC model the soft and hard photon fields are related,  we might expect some $\gamma$-ray - $X$-ray  correlation. Though it is the soft UV photons that are scattered to the GeV energies, and, formally, one expects UV-GeV correlation,  since   $X$-rays and  UV form a continuous spectral distribution, one also expects $X$-ray - GeV correlation as well. Thus we expect  short time-scale statistical correlation between $X$-ray and $\gamma$ rays photons.

\section{Dependence of the $\gamma$-ray luminosity on the spin-down power in the radiation-reaction-limited regime}

Here we discuss the dependence of the $\gamma$-ray luminosity on the spin-down power in the radiation-reaction limit. Generalizing Eq.  (\ref{01}), the total luminosity radiated by a primary beam of Goldreich-Julian density  $n_{GJ}= {B \Omega /( 2 \pi e c)} $ in the radiation reaction limited regime is\be
 L_c = e c \eta B n_{GJ} \eta_G R_{LC}^3
 \label{110}
\ee
where $\eta_G R_{LC}^3$ is the volume occupied by the radiating particles.
Replacing $B$ with a dipole field $B_{NS} * (R_{NS}/R)^3 $ at the light cylinder $R=R_{LC}$,
it follows that
\be
L_c \approx  \eta \eta_G { B_{NS}^2 R_{NS}^6 \Omega^4 \over 2 \pi c^3} \approx  \eta \eta_G \dot{E}_{SD}
\label{11}
\ee
where $\dot{E}_{SD} \approx { B_{NS}^2 R_{NS}^6 \Omega^4 \over 2 \pi c^3}  $,  is the pulsar spin-down power. 

Thus,  {\it  in the radiation-reaction-limited regime, the gamma-ray luminosity is proportional to the  spin-down power},   $L_c \propto \dot{E}_{SD} $. This differs from the   commonly used $L_c \propto \sqrt{ \dot{E}_{SD}}  $ scaling, which results if the maximum particle energy is not limited by radiation reaction but by the electric potential and most of the energy is radiated away once the particle is outside of the accelerating region. This is the case in polar cap models. In these models a beam with a particle density equal to the Goldreich-Julian density loses energy $\dot{N} \propto n_{GJ} r_{PC}^2 \propto \sqrt{\dot{E}_{SD}}$ \citep[see \eg][]{2000ApJ...532.1150Z}, where  $ r_{PC} $ is the radius of the polar cap.   The same  square-root scaling has been extended to outer gaps, assuming that  the emitting volume is proportional to the volume within the  light cylinder radius \citep{2003ApJ...591..334H}. 

The expected linear proportionality (\ref{11}) of the $\gamma$-ray luminosity is valid in the radiation reaction limit, i.e.\,if the dominant radiation processes depend on the particle energy. This is the case, for example,  for curvature radiation or inverse-Compton (IC) scattering in the Thompson regime. And it is not the case for IC scattering  in the Klein-Nishina regime, where the radiative losses are independent of the particle energy (see Eq. \ref{gKN}), and, therefore, the acceleration is not limited by radiation losses. However, as we argued in \S \ref{IC},  there are good reasons to believe that particle acceleration is indeed limited by radiation reaction.

We note that testing our prediction is complicated by the large uncertainty of the geometrical parameter $\eta_G$, the effective emission volume, which depends on the pulsar inclination angle, the angle between the rotation axis, and the line of sight.  It may also depend on the period of the pulsar through the microphysics of the acceleration.

\section{Discussion}

The recent detection of the Crab pulsar above 100\,GeV by VERITAS \citep{VERITASPSRDetection}  changes our picture of high energy gamma-ray emission from pulsars. Even though the breaks in the energy spectra of most pulsars are consistent with curvature radiation in the  radiation reaction limit. For some pulsars exceptional conditions on the accelerating \Efs are required to explain the observed cutoff energies with curvature radiation. In particular, the pulsed emission from the Crab pulsar above 100 GeV can only originate from curvature radiation if extreme assumptions are being made about the pulsar's magnetosphere. The observation that the flux above 100 GeV is in agreement with an extrapolation of the flux from the GeV regime argues in favor of one emission mechanism being dominant below and above the spectral break.
    
 Thus, there are two somewhat independent arguments against curvature radiation as the dominant source of GeV photons: (i) in many pulsars the observed break  energy is too high; (ii) the Crab pulsar energy spectrum above the break is inconsistent with what is expected if the break is due to curvature radiation from particles in the radiation reaction limited regime.  
 
More precise  measurements  of the energy spectra of all $\gamma$-ray pulsars, but especially for those with low break energies, may be decisive for further  progress: for those pulsars for which the break energy is consistent with curvature radiation and moderate electric fields; it is then expected that the energy spectrum above the break follows an exponential cut-off. If, however, the energy spectra above the breaks are better described by power laws like for the Crab, it argues agains curvature radiation. 
 
In this paper we demonstrated  that inverse Compton scattering in the Klein-Nishina regime by the secondary particles results in an overall consistent picture with observations. 
The key features of our model are (i) A population of primaries that is accelerated in a modest  electric field, which is a fraction $\eta$ of the magnetic field strength near the light cylinder with  a typical value of $\eta$ is  $10^{-2}$. The suppression of the scattering cross-section in the Klein-Nishina regime (and the corresponding  lower radiation loss rate of electrons)  allows primary  leptons to be accelerated to very high energies with hard spectra.  (ii) The gain in energy of the primaries in the electric field is balanced by similar curvature radiation and IC losses (radiation reaction limited); (iii)   The secondary plasma is less energetic, but more dense and has approximately the same energy content as the primary beam. The secondary plasma is responsible for the soft UV-$X$-ray emission via synchrotron/cyclotron emission and the high energy
$\gamma$-ray emission that extends to hundreds of GeV via the inverse Compton process. The IC emission from the primary beam extends well into the TeV regime but will be difficult to detected due to the low predicted fluxes. 

Finally, we argued that in the radiation reaction-limited regime the $\gamma$-ray luminosity of pulsars should scale linearly with the spin-down energy, Eq. (\ref{11}). The coefficients of this proportionality depend both on the overall geometry of the \ms\ (\eg\ inclination angle of the magnetic dipole with respect to the axis) through the parameters $\eta_G$ and on the  \Ef\ in the gap through the parameter $\eta$ (which, in turn, depends on 
microphysics of the acceleration precesses). 

This prediction is in contrast to the currently assumed scaling of the $\gamma$-ray luminosity with the available potential,  $\propto \sqrt{\dot{E}_{SD}}$. Observationally, when compared with the scaling of $\propto \sqrt{\dot{E}_{SD}}$, all models underpredict the luminosity of pulsars and thus fail to describe the observed population \cite[\eg][]{2011arXiv1103.2682P} \citep[for alternative interpretation of data see][]{2011ApJ...727..123W}. The proposed linear scaling of the $\gamma$-ray luminosity with the spin down energy naturally predicts more energetic pulsars.

Here we outlined a framework to explain the non-thermal radiation from gaps in the magnetosphere of pulsars. More detailed calculations of the emitted energy spectra are needed. A major complication in including the IC loses in the radiation codes results from the fact that in the KN regime, an accelerating \Ef\ of a given strength does not lead to a fixed energy of a particle. This means that a  particle is either accelerated or decelerated depending on the photon density and the value of the \Ef and does not reach a steady energy. This implies that in this regime acceleration is highly non-stationary. The addition of curvature radiation can, however, establish a steady state. Thus, curvature radiation, even if not dominating the total gamma-ray luminosity, may dictate the particle's final energy. 

A number of additional factors must be taken into account to construct a comprehensive model of the higher energy emission. Most important is the intrinsically  non-isotropic distribution of soft photons. A more detailed structure of the \Bf\ lines within the \ms\ need to be taken into account, including modifications due to magnetospheric currents. Also, particle trajectories may not exactly follow the \Bf\ lines due to  various  drift effects.
 An important modification could be  the IC scattering of the surface thermal emission closer to the surface of the \NS\ in the slot gaps  \citep{1983ApJ...266..215A} \citep[in comparison, Crab does not show any thermal component][]{2004ApJ...601.1050W}.

Our model is based on the assumption that emission is generated within the light cylinder. The main argument for this is that \Fermi\ pulsar profiles are  well fitted with geometric models and that the profile of the  very high energy emission is correlated with the lower energies. This disfavors   the  models that advocate  the emission from the wind zone \citep[\eg][]{2000MNRAS.313..504B,2002A&A...388L..29K}.

We would like to thank  Felix Aharonyan, Jonathan Arons,  Roger Blandford,  Alice Harding, Oleg Kargaltsev,  John Kirk, Mallory Roberts and Roger Romani. 
N. O. was supported in part by a Feodor-Lynen fellowship of the
Alexander von Humboldt Foundation. ML was party supported by  NASA grant NNX09AH37G. 

\bibliographystyle{apj}
 % \bibliography{/Users/maxim/Home/Research/BibTex} \end{document}

\end{document}